\documentclass[prd,floatfix,onecolumn,amsmath,amssymb,floatfix]{revtex4}
\usepackage{graphicx,color,dcolumn,booktabs,bm}
\usepackage{subfigure}
\usepackage{longtable,lscape}
\usepackage{amssymb}
\usepackage{indentfirst}
\usepackage{epsfig}
\usepackage{feynmf}   
\usepackage{epstopdf}   
\usepackage{slashed}  
\usepackage{cases}
\usepackage[pdfpages]{xcolor}
\definecolor{maroon}{RGB}{139,25,150}
\usepackage{multirow}
\usepackage{float}
\usepackage{graphicx,color,dcolumn,booktabs,bm}
\usepackage[colorlinks, citecolor=blue,anchorcolor=red,menucolor=red, linkcolor=red,filecolor=red,runcolor=red,urlcolor=blue,frenchlinks=red, urlcolor=blue]{hyperref}

\begin{document}

\preprint{}
\preprint{}
\title{\color{maroon}{New $\Lambda_b(6072)^0$ state as a $2S$ bottom baryon}}

\author{K.~Azizi}
\affiliation{Department of Physics, University of Tehran, North Karegar Avenue, Tehran
14395-547, Iran}
\affiliation{Department of Physics, Do\u gu\c s University,
Ac{\i}badem-Kad{\i}k\"oy, 34722 Istanbul, Turkey}
\author{Y.~Sarac}
\affiliation{Electrical and Electronics Engineering Department,
Atilim University, 06836 Ankara, Turkey}
\author{H.~Sundu}
\affiliation{Department of Physics, Kocaeli University, 41380 Izmit, Turkey}

\date{\today}

\begin{abstract}
As a result of continuous developments, the recent experimental searches lead to the observations of new particles at different hadronic channels. Among these hadrons are the excited
states of the heavy baryons containing single bottom or charmed quark in their valance quark content.
The recently observed $\Lambda_b(6072)^0$ state is one of  these baryons and possibly $2S$ radial excitation of the $\Lambda_b$ state. Considering this information from the experiment,  we conduct a QCD sum rule analysis on this state and calculate its 
mass and current coupling constant  considering it as a $2S$ radially excited $\Lambda_b$ resonance. For
completeness, in the analyses, we also compute the mass and current coupling constant for the
ground state $\Lambda_b^0$ and its first orbital excitation. We also consider the  $\Lambda_c^+$ counterpart of each state and attain their mass,  as well. The obtained results are consistent with the
experimental data  as well as existing  theoretical predictions.
\end{abstract}

\maketitle
\section{Introduction}

The progress in experimental facilities and techniques culminated in many exciting observations of the various new
particles in recent years. Among these new states, there exist excited states of the heavy baryons at different channels that have been in the focus of much attention. The searches for the properties of these states can  play crucial roles in the understanding of the  dynamics, nature, and quark-gluon organizations of these states as well as the perturbative and nonperturbative natures of QCD. Investigations of the baryons with single heavy quark and two light quarks  contribute to a better understanding of the confinement mechanism and help us test the predictions of not only the quark model and the heavy quark symmetry but also that of other
theoretical approaches used to describe these states.

In the last few decades, we witnessed the observations of various excited baryons containing  single heavy quark in
their quark content. Among these states are the $\Omega_c(3000)^0$, $\Omega_c(3050)^0$, $\Omega_c(3066)^0$, $\Omega_c(3090)^0$, 
$\Omega_c(3119)^0$ states \cite{Aaij:2017nav} observed from the investigation of the $\Xi_c^+ K^-$ mass spectrum, $\Xi_b(6227)^-$~\cite{Aaij:2018yqz}, $\Sigma_b(6097)^{\pm}$~\cite{Aaij:2018tnn}, $\Xi_b'(5935)^-$, $\Xi_b(5955)^-$~\cite{Aaij:2014yka}, $\Lambda_b{}^*(5912)^0$, $\Lambda_b{}^*(5920)^0$~\cite{Aaij:2012da}, $\Lambda_b(6146)^0$, $\Lambda_b(6152)^0$~\cite{Aaij:2019amv}, $\Omega_b(6316)^-$, $\Omega_b(6330)^-$, 
$\Omega_b(6340)^-$ and $\Omega_b(6350)^-$~\cite{Aaij:2020cex}. A wealth of theoretical investigations accompanied these observations to elucidate their various properties and to enrich our understanding of their structures. Their mass spectrum and decay mechanisms were extensively searched for by quark model~\cite{Copley:1979wj,Maltman:1980er,Capstick:1986bm,Ebert:2005xj,Ebert:2007nw,Ebert:2011kk,Garcilazo:2007eh,Valcarce:2008dr,Roberts:2007ni,Karliner:2015ema,Yoshida:2015tia,Shah:2016mig,Shah:2016nxi,Thakkar:2016dna,Ivanov:1998wj,Ivanov:1999bk,Hussain:1999sp,Albertus:2005zy,Migura:2006ep,Zhong:2007gp,Hernandez:2011tx,Liu:2012sj,Chen:2016iyi,Wang:2017kfr,Chen:2018vuc,Wang:2018fjm,Nagahiro:2016nsx,Yao:2018jmc,Wang:2019uaj},
heavy hadron chiral perturbation theory~\cite{Huang:1995ke,Banuls:1999br,Cheng:2006dk,Cheng:2015naa,Jiang:2015xqa}, relativistic 
flux tube model~\cite{Chen:2014nyo}, Bethe-Salpeter formalism~\cite{Guo:2007qu}, $^3P_0$ model~\cite{Chen:2007xf,Ye:2017dra,Ye:2017yvl,Yang:2018lzg,Chen:2017aqm,Guo:2019ytq,Lu:2019rtg,Liang:2019aag}, lattice QCD~\cite{Padmanath:2013bla,Bali:2015lka,Bahtiyar:2015sga,Bahtiyar:2016dom}, the bound state picture~\cite{Chow:1995nw}, light cone QCD sum rules~\cite{Chen:2017sci,Agaev:2017nn,Zhu:1998ih,Wang:2009ic,Wang:2009cd,Aliev:2009jt,Aliev:2010yx,Aliev:2014bma,Aliev:2016xvq,Aliev:2018vye} and QCD sum
rules method~\cite{Zhu:2000py,Wang:2010it,Mao:2015gya,Chen:2016phw,Mao:2017wbz,Wang:2017vtv,Aliev:2018lcs,Cui:2019dzj,Azizi:2020tgh}, etc. For more related discussions about these states, we refer to the Refs.~\cite{Richard:1992uk,Korner:1994nh,Klempt:2009pi,Crede:2013sze,Cheng:2015iom,Chen:2016spr} and the references therein.

Nowadays the LHCb Collaboration announced the observation of another new beauty baryon state, which shows
consistency with $2S$ radial excitation of $\Lambda_b^0$ baryon, in the $\Lambda_b\pi^+\pi^-$ invariant mass spectrum with a significance exceeding 14 standard deviations~\cite{Aaij:2020rkw}. Its mass and width were reported as $m_{\Lambda_b^{**0}}=6072.3\pm 2.9 \pm 0.6 \pm 0.2$~MeV and $\Gamma = 72\pm11\pm 2$~MeV, respectively, with an interpretation of its being $2S$ excited state. This observation is also consistent with the report of CMS collaboration~\cite{Sirunyan:2020gtz} indicating a broad excess of events in the region of $6040-6100$~MeV. In 2012, the LHCb Collaboration announced the observation of two narrow $\Lambda_b$ states decaying into
$\Lambda_b^0\pi^+\pi^-$, which are $\Lambda_b(5912)^0$ and $\Lambda_b(5920)^0$ and these states were interpreted as orbital excitations of $\Lambda_b^0$ baryon~\cite{Aaij:2012da}. These baryons were studied using the QCD sum rule approach in the heavy quark effective theory~\cite{Mao:2015gya}. Later, in 2019, the LHCb collaboration reported the observation of another $\Lambda_b$ baryon doublet, namely $\Lambda_b(6146)^0$ and $\Lambda_b(6152)^0$, with an interpretation of their being $1D$-wave state~\cite{Aaij:2019amv}. The mass predictions in the QCD sum rule method for these states were presented in Refs.~\cite{Chen:2016phw,Azizi:2020tgh}. In the present work, we focus our attention on the newly observed state $\Lambda_b(6072)^0$ and  perform an analysis on the mass of this particle considering its being first radial excitation, $2S$-state, with possible quantum numbers $J^P = \frac{1}{2}^+$, as suggested by the LHCb Collaboration. To this end, we adopt the QCD sum rule method~\cite{Shifman:1978bx,Shifman:1978by,Ioffe81} with a proper interpolating current that couples the states with considered quantum numbers. This method is a non-perturbative method applied with success to calculate various properties of hadrons, such as their spectroscopic and
decay properties, giving consistent results with experimental observations. Thus, the interpolating current used
in the calculations not only couples to the considered radially excited state but also to the ground and orbitally excited
ones. Therefore in this work, we first calculate the mass and the current coupling constant of the ground state baryon,
then we obtain the masses and current coupling constants of its first orbital and radial excitations. For completeness,
we also include in our analyses the charmed counterpart of the considered states. The spectroscopic analyses of the
considered states may shed light on the quantum numbers and structure of these states, improve our understanding of the strong
interaction and help us test the predictions of the quark model.

The outline of this work is as follows: Sec. II provides the details of the QCD sum rules calculations for the masses
and the current coupling constants of the considered states. In Section III the numerical analyses and the results are
presented. The last section gives a summary of the results and conclusion.

\section{QCD sum rule Calculations for the $\Lambda_b$ and $\Lambda_c$ states}

The states considered in this study are analyzed through the following two-point correlation function:
\begin{equation}
\Pi(q)=i\int d^{4}xe^{iq\cdot x}\langle 0|\mathcal{T}\{\eta(x)\bar{\eta}(0)\}|0\rangle ,\label{eq:CorrF1}
\end{equation}        
where $\eta(x)$  represents the interpolating current in terms of  the related valance quark fields and  $\mathcal{T}$ is used to represent the time ordering operator. The following interpolating current is used in the calculations:
\begin{eqnarray}
&&\eta=\nonumber \\ 
&&\frac{\epsilon_{abc}}{\sqrt{6}}\Big[2(u^{aT} C d^b)\gamma_5 Q^c+2\beta(u^{aT}C\gamma_5 d^b)Q^c+(u^{aT}CQ^b)\gamma_5  d^c+\beta(u^{aT} C\gamma_5 Q^b)d^c+(Q^{aT} C d^b)\gamma_5 u^c+\beta(Q^{aT} C\gamma_5 d^b)u^c\Big], \nonumber \\
\label{Eq:cor}
\end{eqnarray}
where $Q$ represents $b$ $(c)$ quark field for  $\Lambda_b$ $(\Lambda_c)$ state; $a$, $b$ and $c$ are color indices, $C$ is the charge conjugation operator and the $\beta$ is an arbitrary parameter to be fixed later. The above interpolating current is written considering all the quantum numbers of the states under study. The three states considered in the present study have all the same quantum numbers and quark contents but different energies, hence, all of these particles couple to the same interpolating field.  According to the quark model, $\Lambda_Q$  belongs to the antitriplet representation of the $SU(3)$ and  the current describing it should be antisymmetric with respect to the exchange of the light quark  fields. The interpolating field should also be  color singlet. Thus, its general form satisfying  these conditions  can be decomposed as
\begin{eqnarray}
\label{kazem}
\eta &\sim & \epsilon_{abc} \Big\{  \Big(
u^{aT} C \Gamma d^b \Big) \tilde{\Gamma} Q^c 
+ \Big( u^{aT} C \Gamma Q^b \Big) \tilde{\Gamma} d^c
+ \Big(Q^{aT} C\Gamma d^b \Big) \tilde{\Gamma} u^c  \Big\},
\end{eqnarray}
where $ \Gamma$ and $ \tilde{\Gamma} $ can be any of the matrices $ 1 $, $ \gamma_5 $, $ \gamma_\mu $, $ \gamma_5 \gamma_\mu$ or $ \sigma_{\mu\nu} $. The main task is  to determine the $ \Gamma$ and $ \tilde{\Gamma}$. For this aim let us first consider the  transpose of the  part $ \epsilon_{abc} (u^{aT} C \Gamma d^b )  $ in the first term:
\begin{eqnarray}
  [\epsilon_{abc} u^{aT} C \Gamma d^b  ]^T=\epsilon_{abc} d^{bT} C (C \Gamma^T C^{-1}) u^a=-\epsilon_{abc} d^{aT} C (C \Gamma^T C^{-1}) u^b,
\end{eqnarray}
where  a simple theorem was used: if $ A=BD $, where $ A $, $ B $ and $ D $ are matrices, whose elements are Grassmann numbers, then $ A^T=-D^TB^T $. We also used  $ C^T=C^{-1} $ and $ C^2 $=-1. The quantity $ C \Gamma^T C^{-1} $ is  $ \Gamma $ for the cases $\Gamma =1 $, $ \gamma_5 $ and  $ \gamma_5 \gamma_\mu$; and it is  
$ -\Gamma $ for the matrices $ \Gamma=\gamma_\mu $  and $ \sigma_{\mu\nu} $.  The transpose of a one by one matrix, i.e., a scalar,  must be equal to itself.  Thus,  
\begin{eqnarray}
-\epsilon_{abc} d^{aT} C (C \Gamma^T C^{-1}) u^b= -\epsilon_{abc} d^{aT} C  \Gamma  u^b= \epsilon_{abc} u^{aT} C \Gamma d^b ,
\end{eqnarray}
which is held for $\Gamma =1 $, $ \gamma_5 $ and $ \gamma_5 \gamma_\mu$. Note that, $ \epsilon_{abc} u^{aT} C \Gamma d^b  $ is antisymmetric for the $ u\leftrightarrow d $  exchange, which was used in the above relation. The simplest way is to choose the  $\Lambda_Q$ state to have the same total spin/spin projection as the heavy quark $ Q $. Therefore,  the spin of the diquark formed by light quarks must be zero. This immediately implies that  $\Gamma =1 $ or $ \gamma_5 $. Thus, the two possible forms of the interpolating  field become
\begin{eqnarray}
\eta_{1} &= & \epsilon_{abc}   \Big(
u^{aT} C  d^b \Big) \tilde{\Gamma}_1 Q^c,
\nonumber\\ 
&\mbox{and}&\nonumber\\ 
\eta_{2} &= & \epsilon_{abc}   \Big(
u^{aT} C\gamma_5  d^b \Big) \tilde{\Gamma}_2 Q^c.
\end{eqnarray}
The matrices $ \tilde{\Gamma}_1 $ and $  \tilde{\Gamma}_2$ are determined via the Lorentz and parity considerations. As $ \eta_{1} $ and $ \eta_{2} $ are Lorentz scalars, one  concludes that $ \tilde{\Gamma}_1 $ and  $  \tilde{\Gamma}_2$   should be $1 $ or $ \gamma_5 $. The parity consideration leads to $ \tilde{\Gamma}_1 =\gamma_5$ and $ \tilde{\Gamma}_2 =1$. Therefore,  the two possible forms of the interpolating field  for the considered term can be written as
\begin{eqnarray}
\eta_{1} &= & \epsilon_{abc}   \Big(
u^{aT} C  d^b \Big) \gamma_5 Q^c,
\nonumber\\ 
&\mbox{and}&\nonumber\\ 
\eta_{2} &= & \epsilon_{abc}   \Big(
u^{aT} C\gamma_5  d^b \Big)  Q^c.
\end{eqnarray}
Evidently, the  arbitrary linear combination of the above possibilities can better represent the baryon  $\Lambda_Q$:
\begin{eqnarray}
\eta &\sim & \epsilon_{abc} \Big[  \Big(
u^{aT} C  d^b \Big) \gamma_5 Q^c 
+ \beta   \Big(
u^{aT} C\gamma_5  d^b \Big)  Q^c\Big],
\end{eqnarray}
where  the general mixing  parameter $ \beta $ is introduced  to  gain the  general form of  the  interpolating field. Repeating  similar steps for the second and third terms in Eq. (\ref{kazem}), we finally acquire   Eq. (\ref{Eq:cor})  to interpolate the  $\Lambda_Q$ states. In the present  study we make an assumption and take the  parameter $ \beta $ the same for all the ground and excited  $\Lambda_Q$  resonances.

According to the standard  prescriptions of the QCD sum rule method, the correlation function is calculated via
two different approaches. First, it is calculated in terms of hadronic degrees of freedom and called  the physical
or hadronic side of the calculations. The result of this part contains the physical quantities such as mass and current
coupling constant of the considered states. The second approach brings out the results in terms of QCD degrees of
freedom such as quark-gluon condensates, QCD coupling constant, the masses of the quarks, etc called the QCD side of the calculations. By matching the results of both sides, considering the coefficients of the
same Lorentz structures, one gets the QCD sum rules for the physical quantities under question.

For the physical side of the calculations, the correlator, Eq.~(\ref{eq:CorrF1}), is calculated by inserting  complete sets of hadronic
states into the appropriate places. This step turns the correlator into the  form
\begin{eqnarray}
\Pi^{\mathrm{Phys}}(q)=\frac{\langle0|\eta|\Lambda_Q(q,s)\rangle\langle\Lambda_Q(q,s)|\bar{\eta}|0\rangle}{m^2-q^2}
+\frac{\langle0|\eta|\tilde{\Lambda}_Q(q,s)\rangle\langle\tilde{\Lambda}_Q(q,s)|\bar{\eta}|0\rangle}{\tilde{m}^2-q^2}
+\frac{\langle0|\eta|\Lambda_Q'(q,s)\rangle\langle\Lambda_Q'(q,s)|\bar{\eta}|0\rangle}{m'{}^2-q^2}+\cdots.
\label{Eq:cor:Phys}
\end{eqnarray}
The $|\Lambda_Q(q,s)\rangle$, $|\tilde{\Lambda}_Q(q,s)\rangle$ and $|\Lambda_Q'(q,s)\rangle$ are used to represent the one-particle states of the ground, and its first orbital excitation $1P$ and first radial excitation $2S$ states, respectively.  Here, $m$, $\tilde{m}$ and $m'$ are their respective masses and $\cdots$ represents the contributions of the  higher states and continuum. The matrix elements  in  Eq.~(\ref{Eq:cor:Phys}) are parameterized as follows:
\begin{eqnarray}
\langle 0|\eta|\Lambda_Q(q,s)\rangle&=&\lambda u(q,s),\nonumber\\
\langle 0|\eta|\tilde{\Lambda}_Q(q,s)\rangle&=&\tilde{\lambda}\gamma_5 u(q,s),\nonumber\\
\langle 0|\eta|\Lambda_Q'(q,s)\rangle&=&\lambda' u(q,s),
\end{eqnarray}  
where $\lambda$, $\tilde{\lambda}$ and $\lambda'$ are the corresponding current coupling constants and $u(q,s)$ is the Dirac spinor. These matrix elements are used in  Eq.~(\ref{Eq:cor:Phys}) and summation over spins of Dirac spinors, which is given as
\begin{eqnarray}
\sum_{s}u(q,s)\bar{u}(q,s)=(\not\!q+m),
\end{eqnarray}  
is applied.  Then, the physical side takes the form:
\begin{eqnarray}
\Pi^{\mathrm{Phys}}(q)=\frac{\lambda^2(\not\!q+m)}{m^2-q^2}+\frac{\tilde{\lambda}^2(\not\!q-\tilde{m})}{\tilde{m}^2-q^2}+\frac{\lambda'^2(\not\!q+m')}{m'{}^2-q^2}+\cdots.
\label{Eq:cor:Phys1}
\end{eqnarray}
After the Borel transformation, the final result for the physical side becomes
\begin{eqnarray}
\tilde{\Pi}^{\mathrm{Phys}}(q)=\lambda^2(\not\!q+m)e^{-\frac{m^2}{M^2}}+\tilde{\lambda}^2(\not\!q-\tilde{m})e^{-\frac{\tilde{m}^2}{M^2}}+\lambda'^2(\not\!q+m')e^{-\frac{m'{}^2}{M^2}}+\cdots.
\label{Eq:cor:Fin}
\end{eqnarray}
%

For the QCD side, one computes the correlation function, Eq.~(\ref{eq:CorrF1}), using the interpolating current given in Eq.~(\ref{Eq:cor})
explicitly. To perform the calculations, first the possible contractions between the quark fields are carried out via
Wick's theorem. For the contracted quark fields the corresponding light and heavy quark propagators presented in
coordinate space are used with following explicit forms:
\begin{eqnarray}
S_{q,}{}_{ab}(x)&=&i\delta _{ab}\frac{\slashed x}{2\pi ^{2}x^{4}}-\delta _{ab}%
\frac{m_{q}}{4\pi ^{2}x^{2}}-\delta _{ab}\frac{\langle \overline{q}q\rangle
}{12} +i\delta _{ab}\frac{\slashed xm_{q}\langle \overline{q}q\rangle }{48}%
-\delta _{ab}\frac{x^{2}}{192}\langle \overline{q}g_{\mathrm{s}}\sigma
Gq\rangle +i\delta _{ab}\frac{x^{2}\slashed xm_{q}}{1152}\langle \overline{q}%
g_{\mathrm{s}}\sigma Gq\rangle  \notag \\
&&-i\frac{g_{\mathrm{s}}G_{ab}^{\alpha \beta }}{32\pi ^{2}x^{2}}\left[ %
\slashed x{\sigma _{\alpha \beta }+\sigma _{\alpha \beta }}\slashed x\right]
-i\delta _{ab}\frac{x^{2}\slashed xg_{\mathrm{s}}^{2}\langle \overline{q}%
q\rangle ^{2}}{7776} ,  \label{Eq:qprop}
\end{eqnarray}%
and
\begin{eqnarray}\label{proheavy} 
S_Q(x) \!\!\! &=& \!\!\! {m_Q^2 \over 4 \pi^2} {K_1(m_Q\sqrt{-x^2}) \over \sqrt{-x^2}} -
i {m_Q^2 \rlap/{x} \over 4 \pi^2 x^2} K_2(m_Q\sqrt{-x^2})
-ig_s \int {d^4k \over (2\pi)^4} e^{-ikx} \int_0^1
du \Bigg[ {\rlap/k+m_Q \over 2 (m_Q^2-k^2)^2} G^{\mu\nu} (ux)
\sigma_{\mu\nu}\nonumber \\
&& +
{u \over m_Q^2-k^2} x_\mu G^{\mu\nu} (ux)\gamma_\nu \Bigg]~,
\end{eqnarray}
where $G_{\mu\nu}$ is the gluon field strength tensor,  $K_{\nu}$ is the Bessel function of the second kind and $G_{ab}^{\alpha\beta}=G_A^{\alpha\beta}t^A_{ab}$, with $A=1,~2,\cdots,8$ and $t^A=\lambda^A/2$. After the usage of the propagators, Fourier and Borel transformations are performed. Finally,   the  continuum subtraction with the help of quark-hadron duality assumption is applied. The result of the QCD side of the sum rule is  obtained in the form
\begin{eqnarray}
\tilde{\Pi}^{\mathrm{QCD}}(s_0,M^2)=\int_{(m_Q+m_u+m_d)^2}^{s_0}dse^{-\frac{s}{M^2}}\rho(s)+\Gamma,
\label{Eq:Cor:QCD}
\end{eqnarray}
where, $s_0$ represents the continuum threshold and  $\rho(s)$ is the spectral density that is obtained by taking the
imaginary part of the result,  $\rho(s)=\frac{1}{\pi}\mathrm{Im}[\Pi^{\mathrm{QCD}}]$.  The   $\rho(s)  $ and $\Gamma  $ are  lengthy  functions, so we don't present their explicit forms here.

After the computations of the both sides, the results are matched through the dispersion relations considering the
coefficients of the same Lorentz structures, that are $\not\!q$  and  $I$.  The QCD sum rules for the considered quantities are
obtained as
\begin{eqnarray}
\lambda^2 e^{-\frac{m^2}{M^2}}+\tilde{\lambda}^2e^{-\frac{\tilde{m}^2}{M^2}}+\lambda'^2e^{-\frac{m'{}^2}{M^2}}=\tilde{\Pi}^{\mathrm{QCD}}_{\not\!q}(s_0,M^2),
\label{Eq:cor:match1}
\end{eqnarray}
 and
\begin{eqnarray}
\lambda^2 m e^{-\frac{m^2}{M^2}}-\tilde{\lambda}^2\tilde{m}e^{-\frac{\tilde{m}^2}{M^2}}+\lambda'^2m'e^{-\frac{m'{}^2}{M^2}}=\tilde{\Pi}^{\mathrm{QCD}}_{I}(s_0,M^2).
\label{Eq:cor:match2}
\end{eqnarray}
 The relation obtained using the $\not\!q$ structure is used to derive the QCD sum rules for
mass and coupling constant by following the  ground state+continuum scheme in which we consider the second and
third terms of the left-hand-side of Eq.~(\ref{Eq:cor:match1}) as parts of the continuum. This results in the following equation for the mass of the  ground state: 
\begin{eqnarray}
m^2=\frac{\frac{d}{d(-\frac{1}{M^2})}\tilde{\Pi}^{\mathrm{QCD}}_{\not\!q}(s_0,M^2)}{\tilde{\Pi}^{\mathrm{QCD}}_{\not\!q}(s_0,M^2)}.
\label{Eq:mass:Groundstate}
\end{eqnarray}
The current coupling constant is obtained as
\begin{eqnarray}
\lambda^2=e^{\frac{m^2}{M^2}}\tilde{\Pi}^{\mathrm{QCD}}_{\not\!q}(s_0,M^2).
\label{Eq:mass:Groundstate}
\end{eqnarray}
 Then
we consider the first two terms on the left-hand side of Eq.~(\ref{Eq:cor:match1}) by increasing the threshold and the third one is taken in the continuum.  By
using the results obtained for ground state as inputs, we get the mass and current coupling constant for the
first orbitally excited, $1P$, state. And finally, the results of the ground  and $1P$ states are used in a similar
manner, namely, ground state+first orbitally excited state+first radially excited state+continuum approach, to obtain the physical
quantities of the radially excited, $2S$, state.

\section{Numerical Analyses}

To numerically analyze the results obtained in the previous section we need  some input parameters that are presented in Table~\ref{tab:Param}.
\begin{table}[tbp]
\begin{tabular}{|c|c|}
\hline\hline
Parameters & Values \\ \hline\hline
$m_{c}$                                     & $1.27\pm 0.02~\mathrm{GeV}$ \cite{Tanabashi2018}\\
$m_{b}$                                     & $4.18^{+0.03}_{-0.02}~\mathrm{GeV}$ \cite{Tanabashi2018}\\
$m_{u}$                                     & $2.16^{+0.49}_{-0.26}~\mathrm{MeV}$ \cite{Tanabashi2018}\\
$m_{d}$                                     & $4.67^{+0.48}_{-0.17}~\mathrm{MeV}$ \cite{Tanabashi2018}\\
$\langle \bar{q}q \rangle (1\mbox{GeV})$    & $(-0.24\pm 0.01)^3$ $\mathrm{GeV}^3$ \cite{Belyaev:1982sa}  \\
$m_{0}^2 $                                  & $(0.8\pm0.1)$ $\mathrm{GeV}^2$ \cite{Belyaev:1982sa}\\
$\langle \frac{\alpha_s}{\pi} G^2 \rangle $ & $(0.012\pm0.004)$ $~\mathrm{GeV}^4 $\cite{Belyaev:1982cd}\\
\hline\hline
\end{tabular}%
\caption{Some input parameters used in the analyses.}
\label{tab:Param}
\end{table}
\begin{figure}[h!]
\begin{center}
\includegraphics[totalheight=6cm,width=8cm]{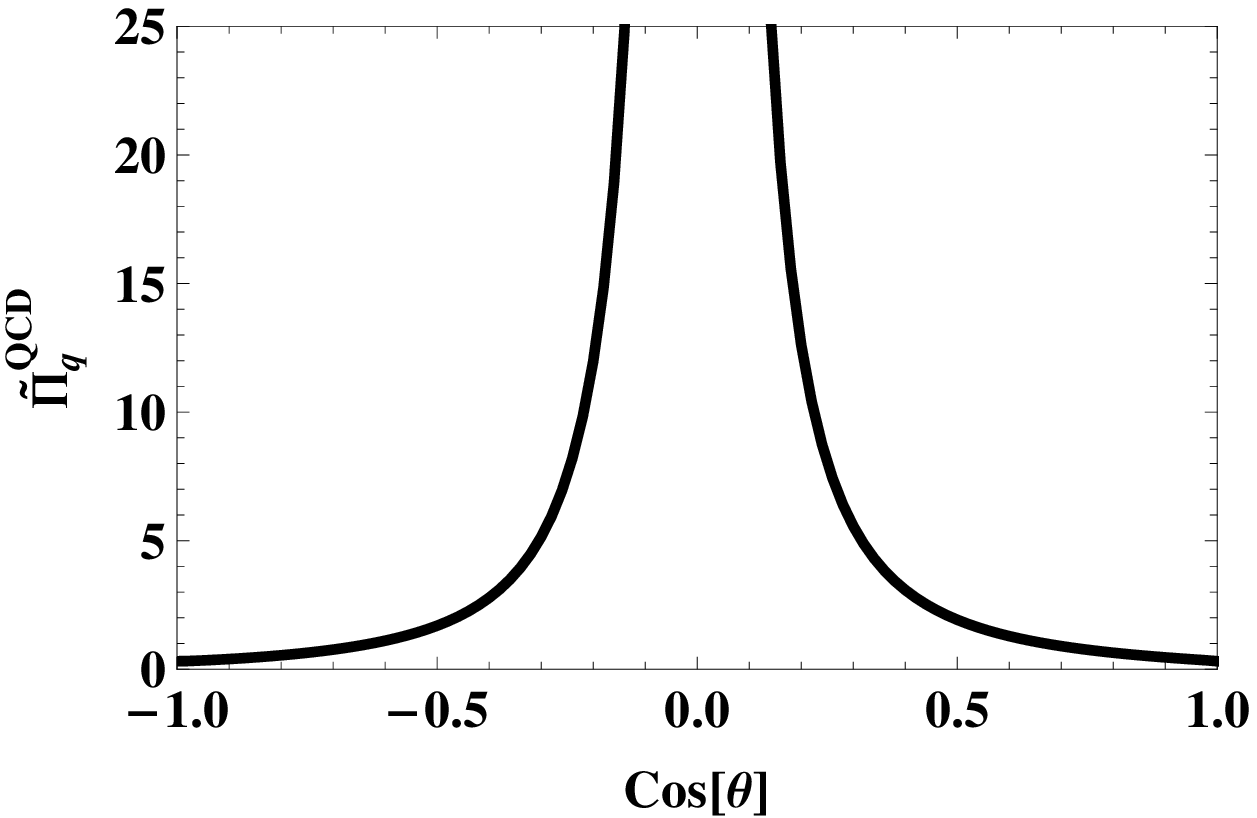}
\end{center}
\caption{  $ \tilde{\Pi}^{\mathrm{QCD}}_{\not\!q} $ as a function of $ \cos\theta $.}
\label{fig:MassT}
\end{figure}
\begin{figure}[h!]
\begin{center}
\includegraphics[totalheight=6cm,width=8cm]{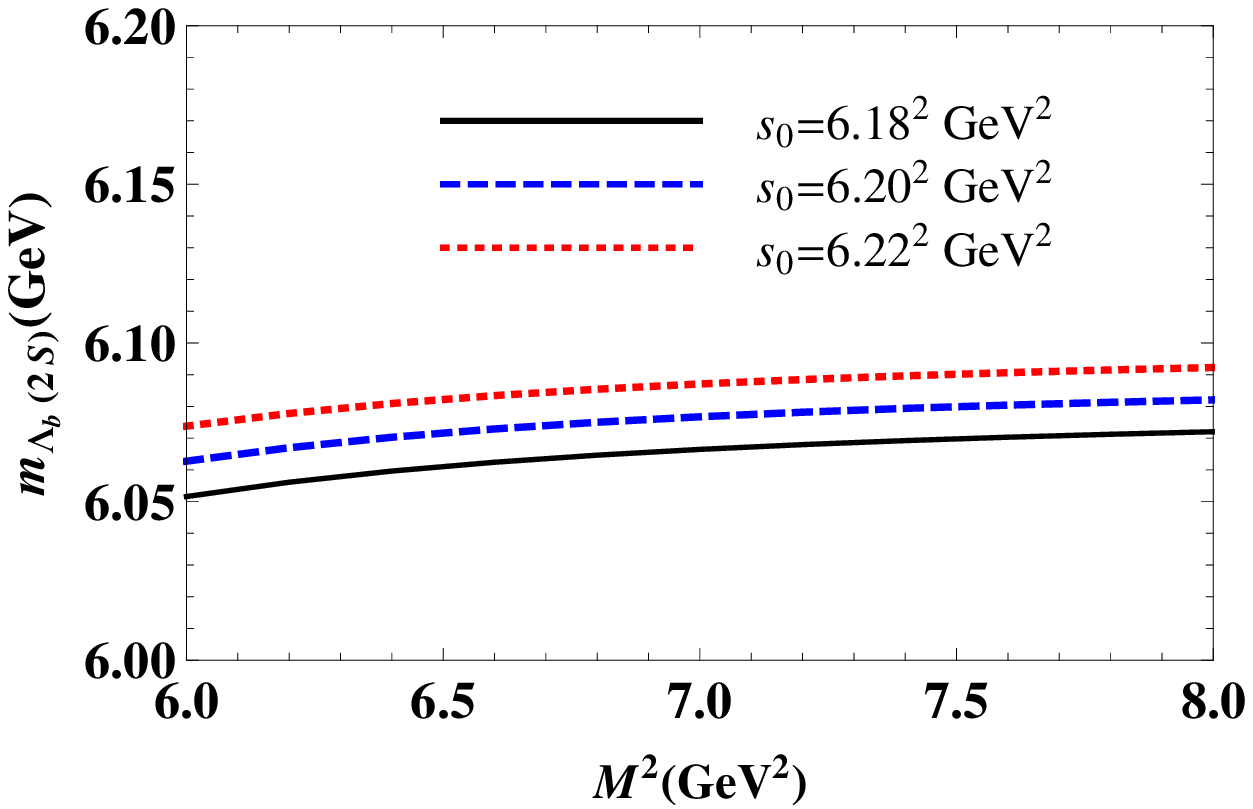}
\end{center}
\caption{  $m_{\Lambda_b} (2S) $ as a function of $M^2  $ at average values of the $ \cos\theta $ and $ s_0 $.}
\label{fig:MassT}
\end{figure}
\begin{figure}[h!]
\begin{center}
\includegraphics[totalheight=6cm,width=8cm]{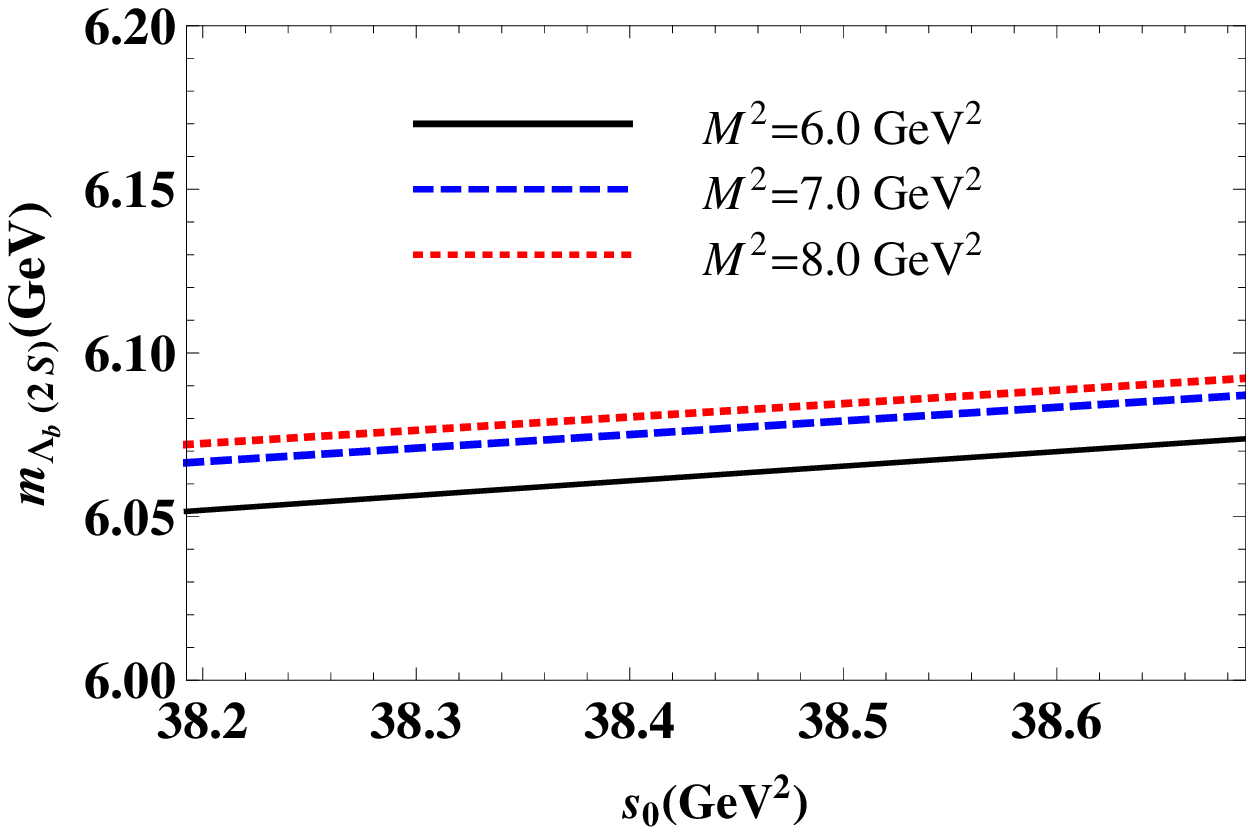}
\end{center}
\caption{  $m_{\Lambda_b} (2S) $ as a function of $s_0  $ at average values of the $ \cos\theta $ and $ M^2 $.}
\label{fig:MassT}
\end{figure}
Though our main concern in the present work is the mass of the newly observed $2S$ $\Lambda_b(6072)^0$ state, we also obtain the
masses for the $1S$ and $1P$ excited states and the corresponding current couplings for both $\Lambda_b^0$ and $\Lambda_c^+$ channels. To this end, we also need to fix three auxiliary parameters that are entered the sum rules: $\beta$,  $s_0$ and $M^2$.  They are fixed based on the standard prescriptions of the method. Thus, we impose the conditions of the mild dependence of the results to the auxiliary parameters, the  convergence of the operator product expansion (OPE), and the dominance of the contributions of the states under consideration over the higher states and continuum.

 To fix the parameter  $\beta$, we plot the functions in the QCD side in terms of this parameter and look for the regions that the results have  weak dependence on the $\beta$. We set $\beta=\tan\theta$ and vary $ \cos\theta $
in the interval $ [-1, 1] $ to explore the whole region. Figure 1, as an example, shows $ \tilde{\Pi}^{\mathrm{QCD}}_{\not\!q} $ as a function of $ \cos\theta $. From this figure and our numerical analyses we obtain the following working windows for $ \cos\theta $, which are valid for all states at $\Lambda_b$ and $\Lambda_c$ 
channels:
\begin{eqnarray}
-1.0 < \cos\theta < -0.5~~~ \mathrm{and}~~~0.5 < \cos\theta < 1.0.
\end{eqnarray}
From figure 1, we see that the $ \tilde{\Pi}^{\mathrm{QCD}}_{\not\!q} $ has a relatively weak dependence on $ \cos\theta $ in the above intervals.

The working intervals of Borel parameters, restricted by the convergence of OPE, the pole dominance requirements, and
the stability of the results in response to the variation of these parameters, are presented in Table~\ref{tab:results}.
For analyses, we take into account the ground-state+first orbitally excited state+first radially excited state+continuum  approach and use Eq.~(\ref{Eq:cor:match1})
\begin{table}[]
\begin{tabular}{|c|c|c|c|c|c|}
\hline
   Particle  & State &$M^2~(\mathrm{GeV^2})$&$s_0~(\mathrm{GeV^2})$  & Mass~(MeV) & $\lambda~(\mathrm{GeV^3})$ \\ \hline\hline
\multirow{3}{*}{} &$\Lambda_b(\frac{1}{2}^+)(1S)$ &$6.0-8.0$& $5.86^2-5.90^2$& $5611.47\pm27.47$ & $0.042\pm0.003$  \\ \cline{2-6} 
   $\Lambda_b$    &$\Lambda_b(\frac{1}{2}^-)(1P)$ &$6.0-8.0$& $5.92^2-5.96^2$& $5910.56\pm84.54$ & $0.020\pm0.008$  \\ \cline{2-6} 
                  &$\Lambda_b(\frac{1}{2}^+)(2S)$ &$6.0-8.0$& $6.18^2-6.22^2$& $6073.65\pm93.22$ & $0.051\pm0.007$ \\ \hline\hline
\multirow{3}{*}{} &$\Lambda_c(\frac{1}{2}^+)(1S)$ &$3.0-5.0$& $2.53^2-2.57^2$& $2282.42\pm28.38$ & $0.022\pm0.001$ \\ \cline{2-6} 
  $\Lambda_c$     &$\Lambda_c(\frac{1}{2}^-)(1P)$ &$3.0-5.0$& $2.63^2-2.67^2$& $2592.36\pm53.01$ & $0.014\pm0.003$ \\ \cline{2-6} 
                  &$\Lambda_c(\frac{1}{2}^+)(2S)$ &$3.0-5.0$& $2.73^2-2.77^2$& $2765.52\pm22.29$ & $0.016\pm0.004$ \\ \hline
\end{tabular}
\caption{The auxiliary parameters and the results of masses and current coupling constants.}
\label{tab:results}
\end{table}
to move step by step as follows: First, we obtain the mass and current coupling constant for the ground state $\Lambda_Q$ particles.
To achieve these quantities we choose proper threshold parameters considering the ground-state+continuum scheme
and the notion that the threshold parameter is related to the energy of the next excited state. Considering that we choose the proper interval for the $s_0$ as also given in Table~\ref{tab:results}. The masses and the current coupling constants obtained
in this step are also given in Table~\ref{tab:results} and these are used as inputs in the second step. Secondly, we consider the ground state+first orbitally excited state+continuum scheme, and with the same logic that is used for the determination of
$s_0$ of the previous step, we determine a new $s_0$ working interval. The results obtained in this step are presented
in Table~\ref{tab:results}, as well. And finally, we consider the  radially excited $2S$ state with ground-state+first orbitally excited state+first radially excited state+continuum approach and attain the proper threshold parameter for this approach. The results obtained for $2S$ states are
also depicted in Table~\ref{tab:results}. The errors in the results arise from the errors of the input parameters and the uncertainties
coming from the determinations of the working intervals for the  auxiliary parameters.   We shall remark that the main source of the uncertainties belongs to the  variations of the parameters $ \beta $, $ s_0 $ and $ M^2 $ in their working windows. Figures 2 and 3 show the dependence of, for instance, $m_{\Lambda_b} (2S) $ to $ M^2 $ and $ s_0 $ at average values of $\beta/cos\theta $. The weak dependence of the mass on $ M^2 $ and $ s_0 $ appears as parts of uncertainties presented in Table~\ref{tab:results}.

\section{Conclusion}

Focusing on the recently observed state $\Lambda_b(6072)^0$, we studied the ground states $1S$, first orbital $1P$ and first radial
$2S$ excitations of the spin-$\frac{1}{2}$ $\Lambda_b$ and $\Lambda_c$ states. The experimentally observed values for the mass of $\Lambda_b(6072)^0$ state is $m_{\Lambda_b^{**0}}=6072.3\pm2.9\pm0.6\pm0.2$~MeV with a width value $\Gamma=72\pm11\pm2$~MeV~\cite{Aaij:2020rkw}. In Ref.~\cite{Aaij:2020rkw}, it was underlined that this result is consistent with the predictions of the quark model for $\Lambda_b(2S)$ state~\cite{Capstick:1986bm,Ebert:2011kk,Roberts:2007ni}. Motivated by this observation,
we calculated the masses and current coupling constants for ground $1S$, first orbitally excited $1P$ and first radially
excited $2S$ states of $\Lambda_b$ and $\Lambda_c$ particles. For the analyses, we applied a powerful nonperturbative method, QCD sum
rule  with a suitable interpolating current formed considering the quark content and quantum numbers of the considered states. The results presented in Table~\ref{tab:results} for ground and first orbital excitations of 
$\Lambda_b$ and $\Lambda_c$ baryons are in good agreement with the present experimental findings given as: $m_{\Lambda_b^0}=5619.60\pm0.17$~MeV~\cite{Tanabashi2018}, $m_{\Lambda_b(5912)^0}=5912.20\pm0.13\pm0.17$~MeV~\cite{Tanabashi2018}, $m_{\Lambda_c^+}=2286.46\pm0.14$~MeV~\cite{Tanabashi2018}, $m_{\Lambda_c(2595)^+}=2592.25\pm0.28$~MeV~\cite{Tanabashi2018}.

As for the main focus of this work, the mass obtained for $\Lambda_b(6072)^0$ as $m_{\Lambda_b(2S)}=6073.65\pm93.22$~MeV is consistent with the experimental result, $m_{\Lambda_b^{0}}=6072.3\pm2.9\pm0.6\pm0.2$~MeV~\cite{Aaij:2020rkw}. The result is also consistent with the various theoretical predictions given for the radially excited $\Lambda_b$ state with $J^P=\frac{1}{2}^+$
as $m=6045$~MeV~\cite{Capstick:1986bm}, $m=6.107$~GeV~\cite{Roberts:2007ni},
$m=6089$~MeV~\cite{Ebert:2011kk}, $m=6106$~MeV~\cite{Valcarce:2008dr}, $m=6153$~MeV\cite{Yoshida:2015tia}. In Ref.~\cite{Thakkar:2016dna} the mass for this particle is calculated using the hypercentral quark model with and without first order corrections to the confinement potential as $m=6.026$~GeV and $m=6.016$~GeV, respectively. The Ref.~\cite{Yang:2017qan} presented the mass of the particle as $m=5982-6127$~MeV obtained from the chiral quark model using five different sets of model parameters. As is seen from these results, the mass obtained in this work is in good consistency with the present theoretical predictions within the errors.

The mass for the $2S$ $\Lambda_c$ state is also obtained for completeness and its value is attained as $m_{\Lambda_c(2S)}=2765.52\pm22.29$~MeV. This result is also consistent with the mass value for $\Lambda_c(2765)^+$ given as 
$m_{\Lambda_c(2765)^+}=2766.6\pm2.4$~MeV~\cite{Tanabashi2018}. This particle is presented in PDG as $\Lambda_c(2765)^+$ or $\Sigma(2765)^+$ with unknown $I(J^P ) =?(??)$ quantum numbers. However in Ref.~\cite{Abdesselam:2019bfp} its isospin was determined as zero and name for it was suggested to be $\Lambda_c(2765)^+$. In this work, we obtained the mass for the first radial excitation of the $\Lambda_c$ state with $J^P=\frac{1}{2}^+$ in consistency with the mass of the $\Lambda_c(2765)^+$
state. Our prediction is also consistent with the theoretical works with the following predictions for $2S$ wave $\Lambda_c$
state: $m=2775$~MeV~\cite{Capstick:1986bm}, $m=2772$~MeV~\cite{Ebert:2007nw}, $m=2769$~MeV~\cite{Ebert:2011kk}, $m=2769$~MeV~\cite{Migura:2006ep}, $m=2.791$~GeV~\cite{Roberts:2007ni}, $m=2772$~MeV~\cite{Chen:2016iyi}, $m=2766$~MeV~\cite{Chen:2014nyo}, $m=2.758$~GeV~\cite{Shah:2016nxi}, $m=2857$~MeV~\cite{Yoshida:2015tia}, $m=2785$~MeV~\cite{Valcarce:2008dr}, $m=2749$~MeV~\cite{Lu:2016ctt} and $m=2654-2825$~MeV~\cite{Yang:2017qan} obtained with five different sets of model parameters. These results are in agreement with that of present work within the errors.

A comparison of the result of this work with the present theoretical and experimental findings indicates that
the particle $\Lambda_b(6072)^0$ is the first radial excitation of the $\Lambda_b$ baryon with the quantum numbers 
$J^P=\frac{1}{2}^+$. The consistency of the result for the first radial excitation of $\Lambda_c$ with $J^P=\frac{1}{2}^+$
with other theoretical results and the present experimental value of $\Lambda_c(2765)^+$ is also considerable. Our result indicates that it may be first radial excitation of $\Lambda_c$ state with quantum numbers $J^P=\frac{1}{2}^+$. Further studies on these states, including their masses and decay properties, and comparison with the result of the present study, may provide more clarifications on the quantum numbers of these states.


\label{sec:Num}

\end{document}